\documentclass[review]{elsarticle}

\usepackage{lineno,hyperref}
\modulolinenumbers[5]

\journal{Journal of \LaTeX\ Templates}

\def\beq{\begin{equation}} 
\def\eeq{\end{equation}}
\usepackage{graphics,epsfig,amssymb}

\begin{document}

\begin{frontmatter}

\title{ Proton-neutron pairing in N=Z  nuclei: quartetting versus pair condensation}

\author[address1]{N. Sandulescu}
\author[address1]{D.Negrea}
\author[address2]{D. Gambacurta}

\address[address1]{National Institute of Physics and Nuclear Engineering, P.O. Box MG-6, Magurele, Bucharest, Romania}
\address[address2]{Istituto Nazionale di Fisica Nucleare - Sezione di Catania, Via S. Sofia 64, I-95123 Catania, Italy}

\begin{abstract}
The isoscalar proton-neutron pairing and  isovector pairing, including both isovector
proton-neutron pairing and like-particle pairing,  are treated in a  formalism which conserves
exactly the  particle number and the isospin.  The formalism is designed for  self-conjugate
(N=Z) systems of nucleons  moving in  axially deformed mean fields and
 interacting through the most general isovector and isoscalar pairing interactions. 
 The ground state of these  systems is described by a superposition of two types of condensates, 
 i.e., condensates of isovector quartets, built by two isovector pairs coupled to the total isospin T=0, 
 and condensates of isoscalar proton-neutron pairs. The comparison with the exact solutions of 
 realistic isovector-isoscalar pairing Hamiltonians shows that this ansatz for the ground state 
 is able to describe with high precision the pairing correlation energies. It is also shown that,
 at variance  with the majority
of  Hartree-Fock-Bogoliubov calculations, in the present formalism the isovector and isoscalar pairing correlations 
coexist  for any pairing interactions. The competition between the isovector and isoscalar
 proton-neutron pairing correlations is studied for N=Z nuclei with the valence nucleons moving in 
 the $sd$ and $pf$  shells and in the major shell above $^{100}$Sn. We find that  in these nuclei the 
 isovector  pairing prevail over the isoscalar pairing, especially for 
 heavier nuclei. However,  the isoscalar proton-neutron correlations  are significant in all nuclei and
 they always coexist with the isovector pairing correlations.
 
\end{abstract}

\end{frontmatter}

%\maketitle

\section{Introduction}

Many nuclei  develop correlations among the valence nucleons which can be treated approximatively as a 
BCS condensate of  Cooper pairs  \cite{bohr}. This approximation works reasonably well  for heavy 
nuclei with neutrons and protons moving in different major shells, in which  the like-particle pairing plays
the dominant role.  However, in spite of many years 
of studies, it is not clear yet which are the physically relevant correlations induced by the pairing interactions
in  nuclei with $N \approx Z$. In particular, the most debated issues are:  (i) whether in $N=Z$ nuclei the pairing can generate 
a condensate of isoscalar proton-neutron pairs; (ii) if this pairing phase  would coexist with the condensate
of isovector proton-neutron pairs and  like-particle pairs; (iii) what could be the fingerprints of a condensate of 
isoscalar proton-neutron pairs in the experimental data (for a recent overview on proton-neutron  pairing
in nuclei see \cite{frauendorf_macchiavelli}). From theoretical point of view the first two issues have been studied mainly in the
framework of Hartree-Fock-Bogoliubov (HFB) approach, which has the advantage of providing an unitary
treatment of like-particle and proton-neutron pairing, both isovector and isoscalar (e.g., see \cite{goodman,bertsch}
and the references quoted therein).  These studies show that: (a) in most of the cases the isovector and isoscalar
proton-neutron pairing correlations  do not coexist; (b) the type of pairing which prevails depends strongly on the
relative strength of isovector and isoscalar  pairing forces.

 In the HFB calculations the particle number and the isospin are not conserved exactly, a drawback which
 could  affect significantly the competition between T=0 and T=1 proton-neutron
 pairing (e.g., see \cite{satula_wyss}). Exactly solvable models in which the particle number and 
 the isospin are conserved  \cite{dobes,engel_so5,dukelsky_so5,engel_so8,lerma_so8} show  that in fact 
 the fundamental ansatz of the HFB theory,  which assumes that the ground state of nuclei can  be described 
 by  a condensate of Cooper pairs, is not appropriate  for N=Z systems.
 Thus, the SO(5) model for isovector pairing interaction shows that in the case of degenerate levels
 the  ground state of N=Z systems is described by a condensate of quartets \cite{dobes} and 
 not by a condensate of Cooper pairs,
 as assumed by the BCS-type approximations. In Refs. \cite{qcm1} it was demonstrated that this is actually the case 
 not only for the schematic SO(5) model but also for any  realistic isovector pairing Hamiltonian.  More precisely, it has 
 been shown that: (i) a condensate of collective quartets, built by two isovector pairs coupled to total isospin T=0,
 describes the pairing correlation energies of N=Z nuclei with a very good precision (errors under 1$\%$); 
 (ii) in nuclei with $N >Z$ the isovector pairing correlations  are accurately described  by a quartet condensate 
 to which it is appended a pair condensate formed by the neutron pairs in  excess \cite{qcm2}; (iii) the isovector pairing, when treated by
  the quartet condensation formalism,  is able to describe reasonably well the Wigner 
  energies in $N \approx Z$ nuclei \cite{qcm_wigner}.

In this paper we extend the quartet formalism of Ref. \cite{qcm1}   for treating both the isovector and the
isoscalar pairing interactions. The  formalism proposed here is dedicated  to those isovector
and isoscalar pairing interactions which  scatter pairs of nucleons in time-reversed states an 
axially-deformed mean fields. These are the pairing interactions which are commonly employed in many 
nuclear structure calculations, e.g., the ones related to beta decays studies \cite{beta_decay}. 

\section{Formalism}

The systems investigated in this study are composed of an equal number of neutrons and protons which
 move in a deformed  mean field with axial symmetry. The nucleons are interacting through
 an isoscalar proton-neutron pairing force and an isovector pairing force, the latter including 
 the proton-neutron pairing and like-particle pairing. The Hamiltonian which describes these systems 
 is given by: 
\beq
\hat{H}= \sum_{i,\tau=\pm 1/2} \varepsilon_{i\tau} N_{i\tau} +
 \sum_{i,j} V^{T=1}(i,j) \sum_{t=-1,0,1} P^+_{i,t} P_{j,t} +
 \sum_{i,j} V^{T=0}(i,j)  D^+_{i,0} D_{j,0} ,\eeq
where $\varepsilon_{i\tau}$ are the single-particle energies associated
to the mean fields of neutrons ($\tau=1/2$) and protons ($\tau=-1/2$). 
In the case of axially-deformed mean fields, supposed here, the single-particle states
are labeled by $i = \{a,\Omega\}$, 
where $\Omega$ is the projection  of the angular momentum on z-axis and
$a$ denotes  the other quantum numbers which specify  the  states.
The second term is the most general isovector pairing interaction expressed by the non-collective
pair operators     
$P^+_{i,1}=\nu^+_i \nu^+_{\bar{i}}$, $P^+_{i,-1}=\pi^+_i \pi^+_{\bar{i}}$ and 
$P^+_{i,0}=(\nu^+_i \pi^+_{\bar{i}} + \pi^+_i \nu^+_{\bar{i}})/\sqrt{2}$.
The third term is the isoscalar proton-neutron pairing interaction and
$D^+_{i,0}=(\nu^+_i \pi^+_{\bar{i}} - \pi^+_i \nu^+_{\bar{i}})/\sqrt{2}$
is the operator which creates a non-collective isoscalar proton-neutron pair.
The operators $\nu^+_i$ and $\pi^+_i$ create, respectively, a neutron and a proton in 
the state $i$ while $\bar{i}=\{a,-\Omega\}$ denotes the time conjugate of the state $i$.

 It can be observed that all pairs operators considered above are constructed with
 the nucleons in time-reversed and axially-deformed states. Therefore the pairs have 
 $J_z=0$, where $J_z$ is the projection of the angular momentum on z-axis, but not 
 a well-defined $J$. In fact,  the isovector pairs and  the isoscalar pairs with $J_z=0$,
 built with axially deformed states, can be seen as a superposition of pairs with $J=\{0,2,4,..\}$ and, 
 respectively, $J=\{1,3,5,..\}$. Therefore the Hamiltonian (1) is not physically equivalent
 with the spherically-symmetric pairing Hamiltonians in which are taken into account  only J=0 isovector
 pairs and J=1 isoscalar proton-neutron pairs. For the latter case a quartet-type formalism, different
 from the one presented below,  has been proposed in Ref. \cite{sasa_plb}.
 
The Hamiltonian (1) has been employed, with various single-particle energies and pairing interactions,
in many studies. In most of them the Hamiltonian (1) was treated in HFB approximation in which, through
a general Bogoliubov transformation, the protons and neutrons are mixed together to form generalized
quasiparticles.  As a consequence, in HFB the particle number and the  isospin are not conserved. 
Here we present a different approach in which both quantities  are conserved exactly from the outset 
through  the way in which the trial wave function is constructed. 

As in Ref.\cite{qcm1}, for describing the isovector  pairing
correlations  we  use as building blocks collective isovector quartets formed by two isovector
 pairs coupled to the total isospin $T=0$, i.e.,
 \beq
A^+ = \sum_{i,j} \bar{x}_{ij} [P^+_i P^+_j]^{T=0} = \sum_{ij} x_{ij}
(P^+_{i,1} P^+_{j,-1}+P^+_{i,-1} P^+_{j,1}
           -P^+_{i,0} P^+_{j,0}).
\eeq
Supposing that the amplitudes $x_{ij}$ are separable in the indices $i$ and $j$, the collective 
quartet  operator can be written as
\beq
A^+= 2 \Gamma^+_1 \Gamma^+_{-1} - (\Gamma^+_0)^2,
\eeq
where $\Gamma^+_{t}= \sum_i x_i P^+_{i,t}$ denote, for t={0,1,-1},  the 
collective Cooper pair operators for the proton-neutron, neutron-neutron and
proton-proton  pairs. 

For treating the isoscalar proton-neutron  correlations we use the collective isoscalar pairs defined by
\beq
 \Delta^+_{0}= \sum_i y_i D^+_{i,0}=\sum_i y_i (\nu^+_i \pi^+_{\bar{i}} - \pi^+_i \nu^+_{\bar{i}})/\sqrt{2} .
 \eeq

With the collective quartet (3) and the collective isoscalar proton-neutron pair (4) we construct the
following approximation for the ground sate of Hamiltonian (1)
\begin{equation}
| \Psi \rangle =(A^+ + (\Delta^+_0)^2)^{n_q} |0 \rangle ,
\end{equation}
where $n_q=(N+Z)/4$ is the number of quartets one can form with the protons and neutrons (N=Z)
participating to the pairing correlations. 

The ansatz (5) for the ground state is suggested by the exact solution of Hamiltonian (1) for  a set of degenerate
states and for pairing forces of equal strength, i.e.,  $g=V^{T=1}(i,j) = V^{T=0}(i,j)$.
We have found that in this case the state (5) is the exact ground state of the Hamiltonian (1). The exact 
ground state energy, when the single-particle energies are put to zero, is given by
\beq
E(n_q,\nu)=2gn_q(\nu-n_q+b) ,
\eeq
where $n_q$ is the number of quartets, $\nu$ is the number of double-degenerate single-particle levels
and $b$=2. It should be noticed that  this particular solution is not the one corresponding to the
isovector-isoscalar pairing Hamiltonian with  SU(4) symmetry \cite{dobes}. In the latter case the 
isoscalar proton-neutron interaction acts in  three channels $\{S=1,S_z= -1,0,1\}$ while here 
we consider only the isoscalar proton-neutron pairs in time-reversed states.

It can be seen that the state (5) is a superposition of terms formed by a product of quartet condensates and
condensates of isoscalar pairs.  In particular, it contains two terms, one formed by a quartet condensate and
the other by a condensate of isoscalar pairs. They are denoted by:
\begin{equation}
| iv \rangle =(A^+ )^{n_q} |0 \rangle ,
\end{equation}
\begin{equation}
| is \rangle =(\Delta^+_0)^{2n_q} |0 \rangle .
\end{equation}
The quartet condensate  (7) is the ansatz used in Refs.\cite{qcm1} to describe the isovector pairing correlations 
in the ground state of N=Z nuclei. From Eq. (3) one can see that the quartet condensate (7) is in fact a superposition
of like-particle and proton-neutron pair condensates. The state (8) is a projected-BCS (PBCS) state, similar to the PBCS
states employed for treating the like-particle pairing. The states (7) and (8)  are the exact solutions of the isovector and, 
respectively, the isoscalar pairing interactions of the Hamiltonian (1) for the case of degenerate states. 
The exact eigenvalues are given by  Eq. (6) with  $b=3/2$ for isovector pairing and $b=1/2$ for isoscalar pairing.
It is interesting to observe that Eq. (6) is in all pairing channels similar to the exact solution of the seniority
model for like-particle pairing (e.g., see \cite{ring_schuck}), the only difference appearing in the value of the quantity $b$. 

The state (5) depends on the parameters $x_i$ and $y_i$ which define the collectivity of isovector and
isoscalar pairs. They are determined variationally from the minimization of the average of the Hamiltonian
and from the condition of normalization of the state (5). To calculate the average of the Hamiltonian on the trial
state (5), preserving  the Pauli principle exactly,  is not a trivial task. In order to evaluate analytically  the
average of the Hamiltonian and the norm  we use the auxiliary states
\beq
| n_1 n_2 n_3  m \rangle = \Gamma_1^{+n_1}  \Gamma_{-1}^{+n_2}\Gamma_0^{+n_3}\Delta_0^{+m}|0\rangle
\eeq
and the recurrence relations method of Ref. \cite{qcm1} . The details of the calculation method, which involves
long expressions , are presented in Ref.\cite{negrea}.

\section{Results and discussions}

One of the most important property of the present formalism for isovector-isoscalar pairing is the prediction that  
all types of pairing correlations coexist  for any pairing interactions. In order to illustrate that, we consider
a system formed by four proton-neutron pairs moving in 10 equidistant levels and interacting through state-independent 
isovector and isoscalar interactions with the strengths given, respectively, by $g_1=g(1-x)/2$  and $g_0=g(1+x)/2$.
For the strength $g$ we take the value 0.6 (in units of the levels spacing) 
while the parameter $x$ is varied between $-1$ and $1$.  In Fig.1 we show how the isovector and isoscalar 
proton-neutron pairing  energies are evolving when one goes from an isovector pairing force to an  
isoscalar pairing force. The proton-neutron pairing energies  are defined as the 
averages $E_{pn}^{T=1}= \langle \Psi \mid g_1\sum_{i,j} P^+_{i,0}P_{j,0} \mid \Psi \rangle $ 
and  $E_{pn}^{T=0}= \langle \Psi \mid g_0\sum_{i,j} D^+_{i,0}D_{j,0} \mid \Psi \rangle $. We observe  that the predictions of the 
present formalism, called hereafter the pair-quartet condensation model (PQCM), follows very closely the exact pairing 
energies (shown by dashed lines) obtained by  diagonalisation. In order to evidence how evolve 
the two types of pairing correlations with the pairing forces, in Fig.1 we display  the overlaps 
between the ground state (5) and the two 
terms of it defined by the quartet condensate (7) and the  condensate of isoscalar pairs (8).  
These overlaps show a  smooth transition from a condensate of quartets to a condensate of pairs, the two types of
correlations coexisting in the ground state  for any ratio between the strengths of the two pairing forces. 
It is worth noticing however that  the relation of these overlaps to the amount of isovector and isoscalar pairing correlations
in the ground state  is not  straightforward  because the state (5) contains, besides the states (7) and (8), a third component 
formed by  the product of the isovector  quartet with the  two isoscalar pairs. Moreover, one should also consider the fact
that  the two states (7) and (8) are not orthogonal to each other (see below).
  Because of these reasons the proton-neutron pairing energies 
 $E_{pn}^{T=0,1}$ have  contributions from both 
 the isovector and the isoscalar degrees of freedom.  Therefore the pairing energies and the  so-called "number of pairs", 
 which are proportional to the former in the case of state-independent pairing forces, cannot be used 
 as  relevant quantities for disentangling the isovector and the isoscalar pairing correlations. 
 
\begin{figure}[t]
\begin{center}
\includegraphics*[scale=0.3,angle=-90]{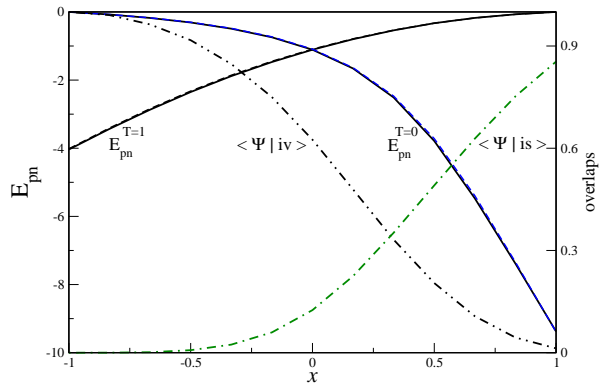}
\caption{Proton-neutron pairing energies provided by the PQCM approach  (full lines) and by 
exact diagonalisation (dashed lines) together with the overlaps between the ground state (5) and 
the states (7) and (8). The parameter $x$ on horizontal axis  scales the strength of the interactions in 
the two pairing channels (see the text). The results correspond to 4 proton-neutron pairs  moving in 10 equidistant
levels and the  pairing energies are given in  units of levels spacing.  }
\end{center}
\end{figure}
 
Next we apply  the present formalism  to analyze the competition between T=1 and T=0 pairing in realistic calculations.
As an example we  consider  N=Z nuclei with the valence nucleons moving outside the closed 
cores $^{16}$O, $^{40}$Ca and $^{100}$Sn. The single-particle states are generated by  Skyrme-HF calculations
performed for axially deformed mean fields.
In the Skyrme-HF calculations, done with the code $ev8$ \cite{ev8}, we use the force $Sly4$ \cite{sly4} and we
disregard the Coulomb interaction.
As the model space for the valence nucleons we consider 10 single-particle levels above the closed cores mentioned above. Since the mean field is axially symmetric, the levels are double degenerate over the projection of the angular momentum on z-axis. In addition, because we neglect the Coulomb interaction, the levels are also degenerate in isospin.

How to fix the pairing interactions in the two pairing channels is a debated issue. Here we shall use the prescriptions  suggested in Refs.\cite{bertsch,bertsch_interaction,gambacurta}. Thus, for the pairing force  we take a zero range delta interaction 
$V^{T=0,1}(\vec{r}_1,\vec{r}_2)=V_0^{T=0,1}\delta(\vec{r}_1-\vec{r}_2)$. The matrix elements of this interaction in the isovector and 
 isoscalar channels  are calculated by projecting out from the two-body wave function the component with the total spin S=0 and, respectively, with $(S=1,S_z=0)$. The strength of the force in  the two channels is  taken as $V_0^{T=1}=V_0$ and $V_0^{T=0}=w V_0$. 
Since the values of the constants $V_0$ and $w$ are also a matter of debate, we have done calculations with various parameters, i.e., $V_0=\{300,465,720\}$ and $w=\{1, 1.25, 1.5, 175\}$. 
Because the conclusions relevant for this study are similar in all these calculations, below we are presenting only the results for
$V_0=465$ and $w=1.5$, which are the values suggested, respectively, in Ref.\cite{bertsch_interaction} and Ref.\cite{bertsch}.

The results of the calculations are displayed in Table I. In the second and third columns are given the 
pairing correlation energies obtained from exact diagonalisation and from PQCM. The correlation energies
are defined as the difference between the total energy and the energy obtained in the absence of the interaction.
One can observe that for all nuclei the agreement between the exact and the PQCM results is excellent. 
Similar good agreements we have obtained for the other pairing forces  mentioned above. 
In the columns 4 and 5 are given the results obtained when the minimization of the Hamiltonian (1) is done 
either with the quartet condensate  (7) or the condensate of isoscalar pairs  (8). It is surprising to see that the
two calculations give results which are not too far from the ones obtained with the full state (5). 
The fact that the calculations with the states (7) and (8)  give comparable results can be understood from 
the overlap $\langle iv \mid is\rangle$  shown in the last column of Table 1. One can thus see that this overlap is 
rather big for all calculated nuclei. From columns (4) and (5) we can notice that for all nuclei the errors corresponding 
to the calculations done with the quartet condensate  (7) are smaller compared to the ones done with the condensate 
of isoscalar pairs (8), indicating that the isovector pairing correlations are stronger than the isoscalar ones, especially
in $pf$-shell nuclei and in the nuclei above $^{100}$Sn . Nonetheless, in all nuclei the isoscalar 
pairing correlations are significant and, as pointed out by the large overlaps shown in column (6), 
they cannot be  disentangled easily from the isovector pairing correlations.

\begin{table}[hbt]
\caption{ Correlation energies calculated in the PQCM approach  compared to the
exact results. Are shown also the correlation energies obtained by minimizing the
Hamiltonian (1) with the isovector $\mid iv \rangle  $  and isoscalar $\mid is \rangle$ states 
defined by Eqs. (7,8). In the last column are given the overlaps between these states. }
\begin{center}
\begin{tabular}{|c|c|c|c|c|c|}
\hline
\hline
   &   exact  & $\mid PQCM  \rangle$ & $\mid iv \rangle$ & $\mid is \rangle$  & $\langle iv \mid  is \rangle$  \\
\hline
\hline
$^{20}$Ne  &  11.38    &  11.38  (0.00\%)  &  11.31  (0.62\%) &  10.92 (4.00\%) &  0.976  \\
$^{24}$Mg  &  19.32    &  19.31  (0.03\%)  & 19.18  ( 0.74\%) & 18.93  (2.00\%) &   0.980 \\
$^{28}$Si  &  18.74   & 18.74     (0.01\%)  & 18.71  ( 0.14\%) &    18.54 (1.07\%)  & 0.992   \\
\hline
$^{44}$Ti  &  7.095   & 7.094  (0.02\%)  & 7.08  (0.18\%) & 6.30  (10.78\%) & 0.928  \\
$^{48}$Cr  &  12.78    & 12.76  (0.1\%)  & 12.69  ( 0.67\%) & 12.22  (4.37\%) & 0.936  \\
$^{52}$Fe  &  16.39    & 16.34  (0.26\%)  & 16.19  ( 1.17\%) & 15.62 (4.65\%)   & 0.946  \\
\hline
$^{104}$Te & 4.53    & 4.52 (0.06\%)  & 4.49 (0.82\%) &  4.02 (11.26\%) & 0.955 \\
$^{108}$Xe & 8.08     & 8.03  (0.61\%)  & 7.96 (1.45\%) & 6.75 (16.47\%)   & 0.814  \\
$^{112}$Ba &  9.36    & 9.27  (0.93\%)  & 9.22  (1.43 \%) &  7.50 (19.81\%)  & 0.784 \\
\hline
\hline
\end{tabular}
\end{center}
\end{table}

Finally we would like to mention that  the main conclusion of this study, namely the coexistence of the
isovector and isoscalar pairing correlations for  any N=Z nuclei, refers to  pairing forces acting on time-reversed
and axially-deformed states. It is however worth mentioning that a similar conclusion was found recently for
spherically-symmetric Hamiltonians with J=0 and J=1 pairing forces in which all the components  of the 
isoscalar J=1 pairing force have been taking into account, not only the one  scattering pairs in time-reversed 
states \cite{sasa_plb}.  We recall that in Ref. \cite{sasa_plb} the isoscalar J=1 pairing is treated by isoscalar
quartets built  by two J=1 pairs coupled to the total angular momentum J=0.  This formalism cannot be 
applied for the isoscalar pairing interactions acting on  deformed states considered in this study since in this case 
the pairs  have not a well-defined angular momentum.

\section{Summary}

In this paper we have proposed a new approach for treating the isovector and the isoscalar
pairing interactions in axially-deformed N=Z nuclei.
In this approach, which conserves 
exactly the particle number and the isospin,  the ground state is constructed as a superposition of condensates
formed by isovector quartets and isoscalar pairs. It is shown that this ansatz for the ground state 
is able to provide very accurate pairing correlation energies 
for all N=Z nuclei analysed in this study. One of the important predictions of this formalism is that the 
isovector and the isoscalar correlations coexist for any pairing interaction. 
In   addition, the realistic calculations presented in this study indicate that the isovector and the isoscalar
correlations are strongly mixed together and difficult to disentangle from each other. 

\vskip 0.4cm
\noindent
{\bf Acknowledgements}
\vskip 0.2cm
\noindent
This work was supported by the Romanian Ministry of Education  and Research through the 
grant Idei nr 57.

%\section*{References} 

\bibliography{biblio}

\end{document}